\documentclass[twocolumn,superscriptaddress,showpacs,preprintnumbers,amsmath,amssymb,pre]{revtex4}
\usepackage{amsmath}
\usepackage{amssymb}
\usepackage{dcolumn}
\usepackage{graphicx}
\usepackage{bm}
\usepackage{float}

\begin{document}
\title{Experiments demonstrate that the null space of the rigidity matrix 
determines grain motion during vibration-induced compaction} 
\author{Aline Hubard}
\affiliation{Department of Physics and Benjamin Levich Institute, The City College of the City University of New York, New York, 10031, USA}
\author{Corey S. O'Hern}
\affiliation{Department of Mechanical Engineering and Materials Science, Yale University, New Haven, Connecticut, 06520, USA}
\affiliation{Department of Physics, Yale University, New Haven, Connecticut, 06520, USA}
\affiliation{Department of Applied Physics, Yale University, New Haven, Connecticut, 06520, USA}
\author{Mark D. Shattuck}
\affiliation{Department of Physics and Benjamin Levich Institute, The City College of the City University of New York, New York, 10031, USA}
\affiliation{Department of Mechanical Engineering and Materials Science, Yale University, New Haven, Connecticut, 06520, USA}

\date{\today}

\begin{abstract}
Using a previously developed experimental method to reduce friction in
mechanically stable packings of disks, we find that frictional
packings form tree-like structures of geometrical families that lie on
reduced dimensional manifolds in configuration space.  Each branch of
the tree begins at a point in configuration space with an isostatic
number of contacts and spreads out to sequentially higher dimensional
manifolds as the number of contacts are reduced.  We find that
gravitational deposition of disks produces an initially
under-coordinated packing stabilized by friction on a high-dimensional
manifold. Using short vibration bursts to reduce friction, we compact
the system through many stable configurations with increasing contact
number and decreasing dimensionality until the system reaches an
isostatic frictionless state. We find that this progression can be
understood as the system moving through the null-space of the rigidity
matrix defined by the interparticle contact network in the direction
of the gravitational force.  We suggest that this formalism can also
be used to explain the evolution of frictional packings under other
forcing conditions.
\end{abstract}

 \maketitle

One of the great successes of statistical mechanics is the ability to
calculate the average physical properties of macroscopic systems in
thermal equilibrium, e.g. atomic and molecular liquids.  The
microstates and probabilities with which they occur are required for
determining the average properties. For example, in the microcanonical
ensemble, each of the microstates at fixed energy is equally probable.

Here, we are interested in determining the structural and mechanical
properties of static packings of frictional granular materials, which
are collections of macroscale grains that interact via repulsive contact
forces~\cite{rmp}. Our studies focus on {\it frictional} packings, which 
have not received as much attention as idealized packings of 
frictionless particles~\cite{liu}. Granular packings are also strongly out-of-equilibrium
since thermal fluctuations do not give rise to particle
rearrangements.  Grain motion is induced instead through applied
external loads~\cite{nicolas}. Moreover, the current state of the
system may depend strongly on the protocol used to generate
it~\cite{protocol1,protocol2,protocol3}. In these systems, the
definition of the appropriate microstates, their probabilities, and
the framework for calculating average quantities is still under
development~\cite{edwards,henkes,makse}.

In previous numerical simulations~\cite{gao2} and experimental
studies~\cite{gao}, we showed that static packings of {\it
frictionless} disks occur as distinct points in configuration space,
whose number grows with the number of particles $N$ in the system.  We
characterized each static packing using an invariant of the
interparticle contact matrix and showed that static packings were not
sampled with equal probabilities when generated using typical
isotropic compression algorithms~\cite{gao2}.

In recent simulations, we showed that static packings of {\it
frictional} disks do not occur as points in configuration space, but
instead as lines, areas, and higher dimensional structures ({\it i.e.},
geometrical families~\cite{shen}) with a dimension $m$ that grows with
the difference in the number of contacts of the system from the
isostatic value~\cite{isostatic} for frictionless disks, $m=N_c^{\rm iso} -
N_c >0$~\cite{shen}. Thus, at each $m$, there are an infinite number of
packings of frictional disks, but they can be classified into a finite
set of geometrical families with the same contact network.

Granular systems can undergo rearrangements from one static packing to
another in the presence of an applied load, {\it e.g.}, slow
compaction during vertical tapping~\cite{vibration}.  There have been
a number of theoretical studies of slow compaction in vibrated
granular packings including Tetris-like and other lattice
models~\cite{nicodemi} and adsorption-desorption or parking-lot
models~\cite{talbot}, which have provided phenomenological
understanding of the logarithmic relaxation of the density. Here, we
seek a particle-level understanding of rearrangements that occur
during vibration in terms of the motion along geometrical families and
from one geometrical family to another.

In this Letter, we describe quasi-two dimensional experiments of
vibrated stainless steel particles and detailed analyses of the
vibration-induced particle motion. We successively applied small
amplitude, high frequency vibration bursts to the packings to break
interparticle contacts and induce small-scale particle rearrangements.
Using accurate particle tracking techniques, we monitored changes in
the interparticle contact network as the system evolved from dilute to
more dense packings during the vibrations. We found two key results:
1) The packings that occur during the compaction process are not
randomly distributed in configuration space, instead they form
connected lines or `geometrical family trees', which verifies our
previous simulation results~\cite{shen}. 2) The motion of the grains
during the compaction process occurs in the null space of an 
under-coordinated rigid link network~\cite{thorpe} formed by the
interparticle contacts.  In particular, the particle motion can be
accurately predicted by projecting the direction of the gravitational
force onto the null space. 

\begin{figure}
\includegraphics[width=2.8in]{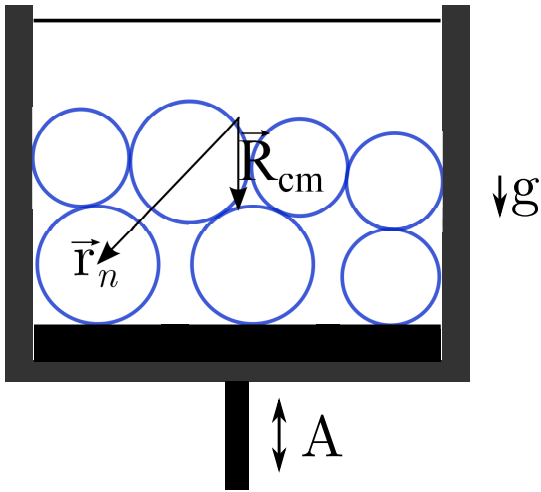}
\caption{(Color online) Sketch of the quasi-two dimensional vibration
cell that contains a packing of seven bidisperse thin cylinders ($4$
small and $3$ large with diameter ratio $1.246$) with $11$ contacts.
The amplitude of the vertical vibrations is indicated by $A$ and
gravity is directed downward. ${\vec R}_{\rm cm}$ and ${\vec r}_n$
locate the center of mass of the packing and position of the $n$th
particle.}
\label{fig:one}
\end{figure}

We performed quasi-two dimensional experiments of vibrated granular
materials as shown in Fig.~\ref{fig:one}. The vibration cell has width
$L_x = 5.33~{\rm cm}$, height $L_y > 5~{\rm cm}$, and thickness
$L_z=0.34~{\rm cm}$ and contained $N=7$ bidisperse stainless steel
cylinders ($4$ small and $3$ large with diameters $\sigma_s =1.26~{\rm
cm}$ and $\sigma_l=1.57~{\rm cm}$, masses $m_s=2.8~{\rm g}$ and
$m_l=4.3~{\rm g}$, and thickness $0.32~{\rm cm}$).  We focused on
systems with a small number of grains so that we can enumerate most of
the geometrical family trees during compaction. The vibration cell is
formed by two plastic side walls, two glass plates, and a movable
piston on the bottom. The diameter of each cylinder is a factor of $4$
larger than its thickness, which ensures that the cylinder axes are
always parallel to each other and perpendicular to the direction of
gravity. The piston is connected to a computer controlled
electromagnetic shaker that vibrates the bottom boundary.

For the experimental protocol, we first randomize the initial
positions of the cylinders by applying high amplitude and low
frequency ($50~{\rm Hz}$) vibrations for one second and allow them to
settle to a frictionally stabilized packing.  We then periodically
apply small amplitude, high frequency ($440~{\rm Hz}$) vibration
bursts, each with duration $10~{\rm ms}$. The acceleration from these
bursts causes the contacts between the cylinders (and between the
cylinders and walls) to break, which temporarily removes the
frictional forces between particles. After each burst, the system
settles under gravity, contacts reform, and the system forms a new
static packing.  Over a series of bursts the system compacts and the
number of contacts grows on average, terminating on a `frictionless'
packing with $N_c = N_c^{\rm iso}=2N=14$ contacts for systems with
fixed boundaries and gravity. We analyzed $6901$ initial packings
without rattler particles on the bottom boundary and applied $20$
vibration bursts to each.  After roughly $15$ bursts, these systems
converged to $979$ distinct frictionless packings.

We imaged the system with a digital camera using backlighting. 
We recorded images of the system after waiting $0.5~{\rm s}$ following 
each burst.  We tracked the positions of the cylinders with a
spatial resolution of $6\times 10^{-6} \sigma_s$ in both the
horizontal and vertical directions~\cite{book}.

\begin{figure}
\includegraphics[width=3.2in]{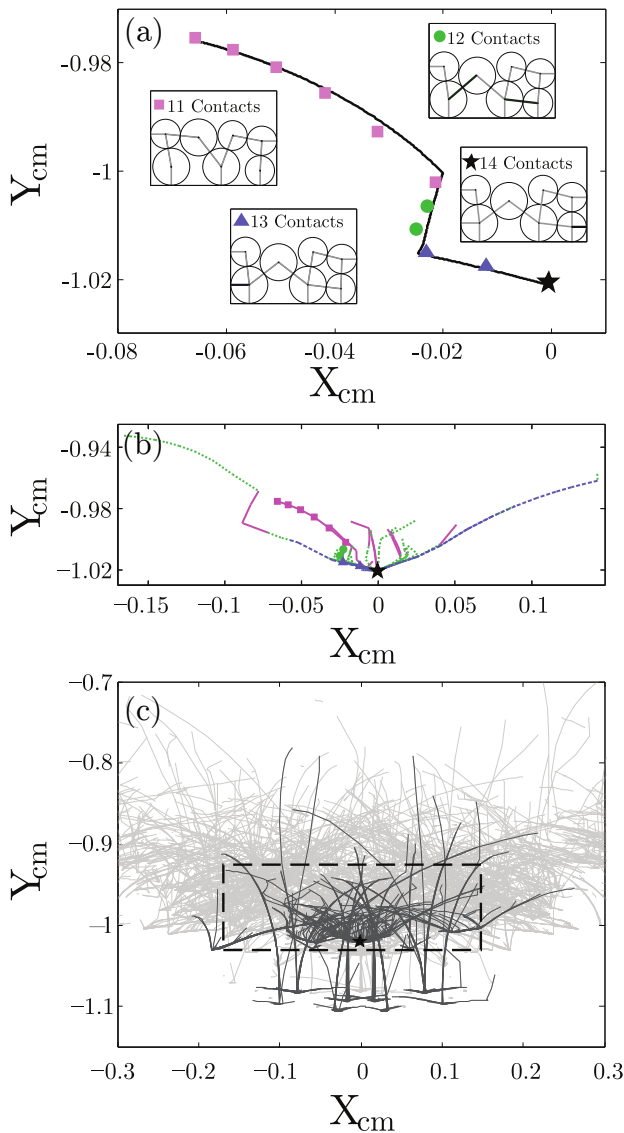}
\caption{(Color online) (a) Evolution of the center of mass ($X_{\rm
cm}$,$Y_{\rm cm}$) as the system compacts and the contact number
increases from $N_c=11$ to $14$.  The star indicates a frictionless
packing with $N_c^{\rm iso}=14$ contacts.  Snapshots of this
`geometrical family tree' are included for each value of $N_c$. The
solid line shows the solution of Eq.~\ref{newton} in the overdamped
limit for each of the contact networks. (b) Evolution of the centers
of mass for all family trees (different line types) that terminate on
the frictionless packing in (a). (c) All family trees found in the
vibration experiments.  The family trees with probability greater than
$0.6\%$ are shaded black, while the others appear gray.  The dashed
rectangular region indicates the area shown in (b).}
\label{fig:two}
\end{figure}

In Fig.~\ref{fig:two} (a), we show the evolution of the center of mass
${\vec R}_{\rm cm}$ and contact number for a set of packings that compacts
to a particular frictionless packing with $N_c^{\rm iso}$
contacts. Mechanically stable frictional packings with fixed
boundaries and gravity can occur in the range $(3N-1)/2 \le N_c \le
2N$~\cite{vanhecke,stefanos} or $10 \le N_c \le 14$ for $N=7$. The
system is initialized in a dilute packing with $N_c=11 < N_c^{\rm
iso}$.  The center of mass evolves smoothly until the packing becomes
mechanically unstable, and a new contact network is formed, in this
case with $N_c=12$. The $N_c=12$ contact network smoothly evolves
until it becomes unstable, and a new interparticle contact forms with
$N_c=13$.  The evolution of the center of mass along a geometrical
family is continuous as long as the contact network remains the same.
Changes in the contact network are signaled by abrupt changes in the
direction of motion of the center of mass. Since we apply small
amplitude vibrations, the evolution of the center of mass terminates
on a frictionless packing with $N_c^{\rm iso}$ contacts.

When the system is initialized in a different packing, it can evolve
to the same frictionless packing in Fig.~\ref{fig:two} (a) or another
one.  In Fig.~\ref{fig:two} (b), we show the geometrical family tree
that leads to the frictionless packing shown in Fig.~\ref{fig:two}
(a). Note that the center of mass approaches that of the frictionless
packing only along particular directions in configuration space. In
Fig.~\ref{fig:two} (c), we show the centers of mass of all packings
found in the experiments. Note that the most probable packings tend to
be the most compact.

We now describe a method to quantitatively predict the particle motion
that occurs during the compaction process. The center of mass motion
is characterized by smooth evolution while the contact network is
fixed, punctuated by abrupt changes in direction when the contact
network changes.  While the contact network is fixed, we model the
system by a network of rigid links between contacting particles.
Since $N_c^{\rm iso} - N_c = m \ge 0$, the $2N\times 2N$ rigidity
matrix $K$ will possess $m$ floppy eigenmodes ${\bf
e}^i=\{e_{x1}^i,e_{y1}^i,\ldots,e_{xN}^i,e_{yN}^i\}$ that form the
null space and $2N-m$ non-floppy modes~\cite{thorpe}. Each contact
network will possess a different rigidity matrix and set of null-space
modes. We define the rigidity matrix and its eigenmodes ${\bf
e}^i_{s,n}$ for each experimental packing, where the subscripts $s$
and $n$ indicate the initial condition and vibration burst number,
respectively.  We have shown previously that vibrations release the
frictional contacts~\cite{gao}.  We now hypothesize that the release
of friction induces motion in the null space along directions that do
not cost energy, and thus frictional packings live in the null space.

\begin{figure}
\vspace{0.1in}
\includegraphics[width=2.8in]{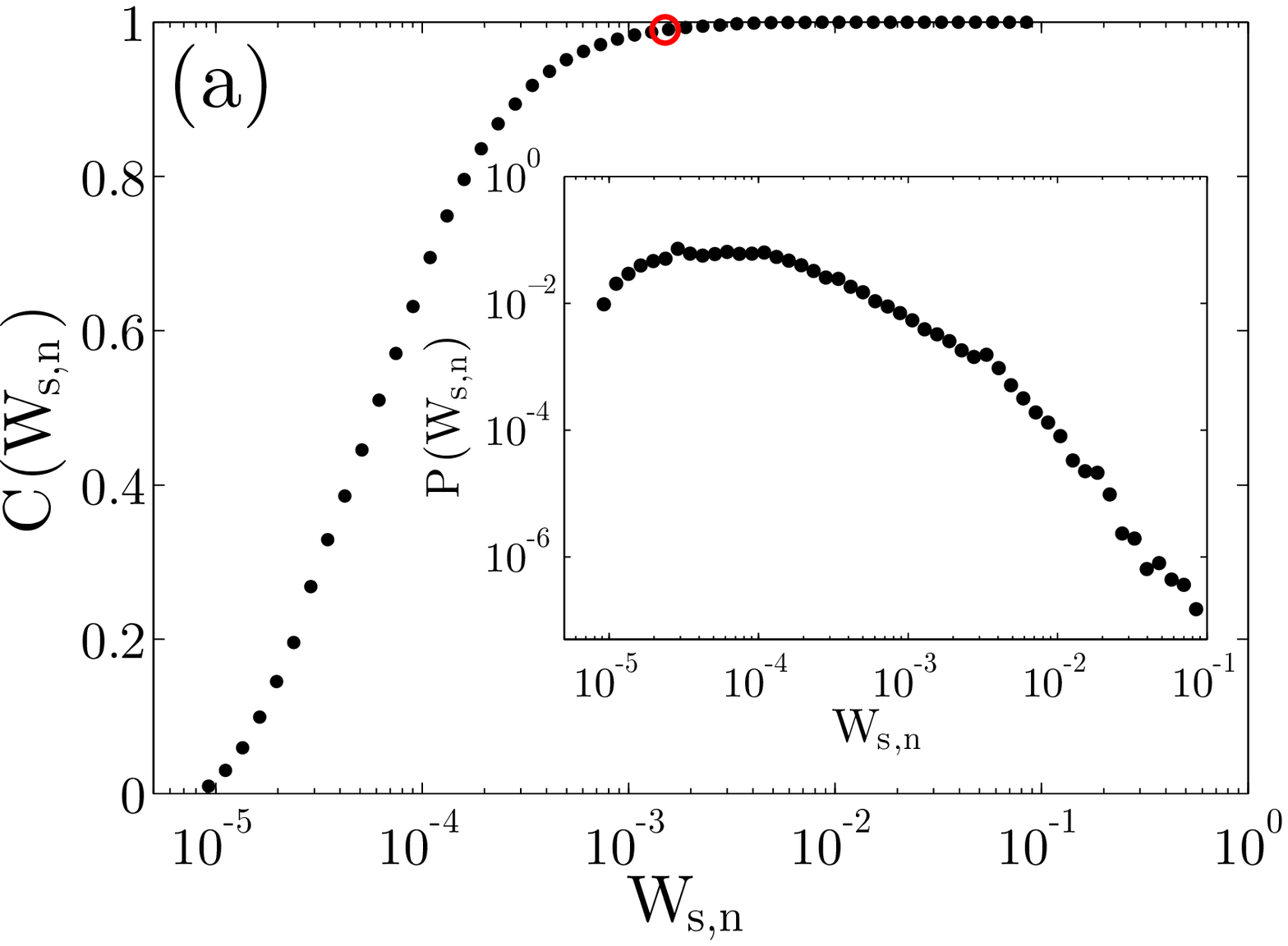}
\includegraphics[width=2.8in]{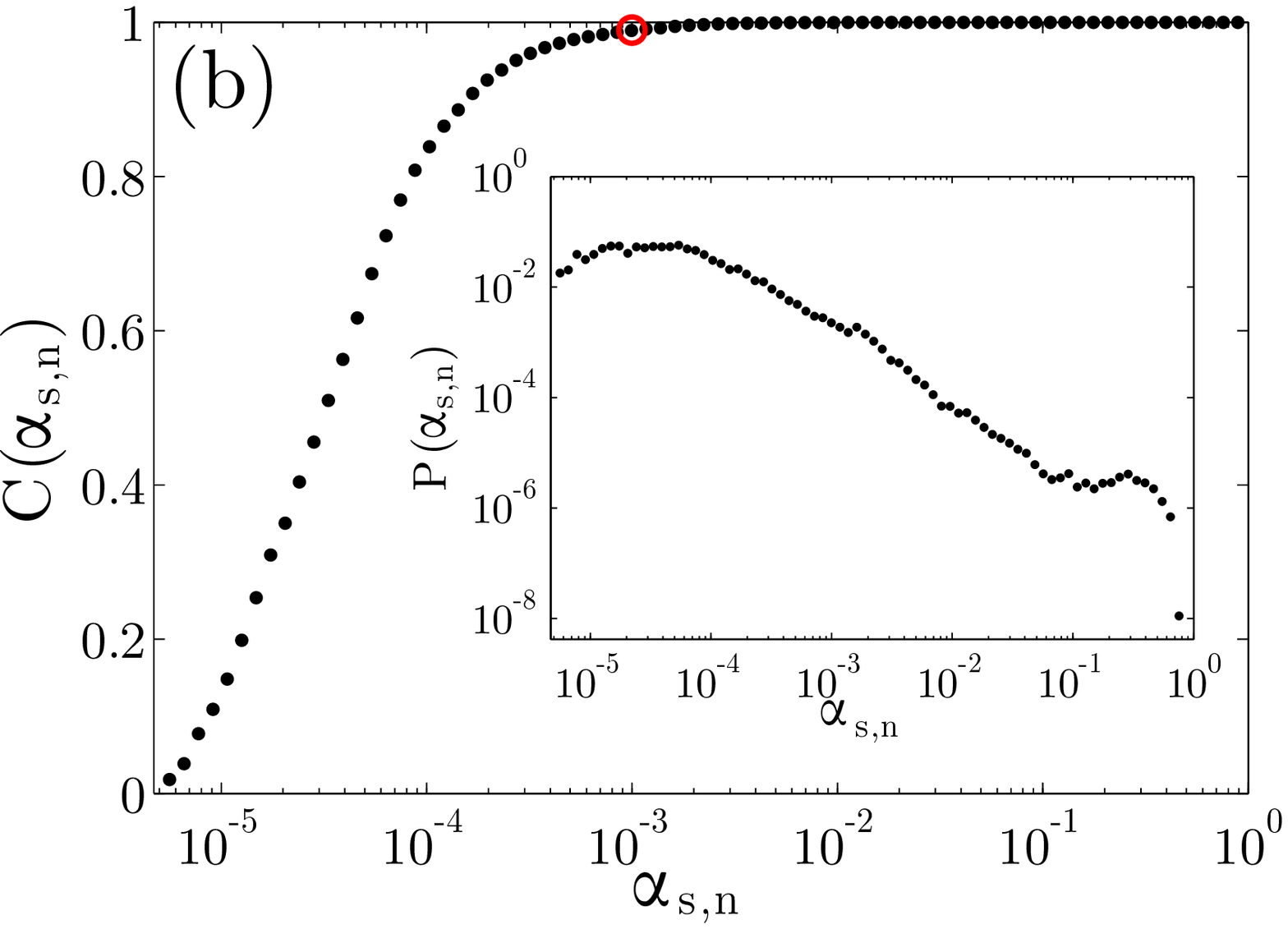}
\caption{(Color online) (a) Cumulative distribution $C({\rm W}_{s,n})$ of
the normalized differences ${\rm W}_{s,n}$ between the particle motion and
the motion that occurs in the null space for each initial condition
$s$ and vibration burst $n$. (See Eq.~\ref{null}.) The open red circle
indicates that $99\%$ of the projections satisfy ${\rm W}_{s,n} \lesssim
10^{-3}$. The inset shows the probability distribution
$P({\rm W}_{s,n})$. The minimum ${\rm W}_{s,n} \sim 10^{-5}$ is set by the
resolution of the particle tracking. (b) Cumulative distribution
$C(\alpha_{s,n})$ of the deviations from complete overlap
$\alpha_{s,n}$ between the particle displacements and direction of
gravity for each $s$ and $n$. The open red circle indicates that
$99\%$ of the overlaps satisfy $\alpha_{s,n} \lesssim 10^{-3}$.  The
inset shows the probability distribution $P(\alpha_{s,n})$.}
\label{fig:three}
\end{figure}

To test whether the particle motion is confined to the null space, 
we calculate the normalized difference between the total particle motion and 
the motion that occurs in the null space,
\begin{equation}
\label{null}
{\rm W}_{s,n} = \frac{ (\Delta {\bf r}_{s,n})^2 - \sum_{i=1}^m \left( \Delta {\bf r}_{s,n} \cdot 
{\bf e}^i_{s,n} \right)^2}{(\Delta {\bf r}_{s,n})^2},
\end{equation}
where $\Delta {\bf r}_{s,n} = {\bf r}_{s,n+1}-{\bf r}_{s,n}$ and ${\bf
r}_{s,n}$ is the $2N$-dimensional vector that gives the positions of
the particles for initial condition $s$ and burst $n$. ${\rm W}_{s,n}
= 0$ indicates that all particle motion is confined to the null space.
In Fig.~\ref{fig:three} (a), we plot the cumulative distribution
$C({\rm W}_{s,n})$, which shows that $99\%$ of the particle motions
satisfy ${\rm W}_{s,n} \lesssim 10^{-3}$. Thus, nearly all of the particle
motion during compaction is confined to the null space.
 
To complete our theoretical analysis of vibration-induced compaction,
we further hypothesize that particle motion occurs in the direction of
gravity projected onto the null space.  We now compare the direction
of motion in configuration space during compaction to the direction of
the gravitational force decomposed onto to the null space,
\begin{equation}
\label{gravity}
{\bf G}_{s,n} = \sum_{i=1}^m \left[ {\bf e}^i_{s,n} \cdot (-{\bf g}) \right]
{\bf e}^i_{s,n},
\end{equation}
where ${\bf g} = \{0,m_sg,\ldots,0,m_lg\}$ is a $2N$-dimensional
vector with a zero for the $x$-component and the gravitational force
for the $y$-component for each particle.  The degree to which the
particle displacements track the direction of gravity can be
determined by calculating the normalized overlap between ${\bf
G}_{s,n}$ and $\Delta {\bf r}_{s,n}$. To measure the deviation from complete 
overlap, we define
\begin{equation}
\label{alpha}
\alpha_{s,n} = 1 - \frac{{\bf G}_{s,n} \cdot \Delta {\bf r}_{s,n}}{|{\bf G}_{s,n}| |\Delta {\bf r}_{s,n}|}.
\end{equation}
$\alpha_{s,n}=0$ indicates that a given displacement $\Delta {\bf
r}_{s,n}$ is aligned with the direction of gravity in the null
space. In Fig.~\ref{fig:three} (b), we plot the cumulative
distribution $C(\alpha_{s,n})$, which shows that $99\%$ of the
particle motions satisfy $\alpha_{s,n} \lesssim 10^{-3}$.  Thus,
nearly all of the particle displacements are in the direction of
gravity in the null space.  However, a few displacements out of
$10^{5}$ possess $\alpha_{s,n} \sim 1$ (with overlap angles greater
than $30^\circ$) indicated by the small peak in the probability
distribution $P(\alpha_{s,n})$ in the inset to Fig.~\ref{fig:three}
(b). These few instances of large $\alpha_{s,n}$ are caused by two
effects: 1) The vibration bursts were not large enough to break some
of the frictional contacts between particles and the bottom wall. In
which case, these particles did not move in the direction of gravity
in the null space. These occurrences of large $\alpha_{s,n}$ can be
removed by increasing the amplitude of the vibration or by adding
constraints to the rigidity matrix to represent frictional
contacts. 2) The particle positions in the packing are such that
different subsequent particle motions are equally likely. For example,
when a particle falls vertically downward on top of another particle.
Whether the particle moves downward and to the right or to the left is
sensitive to the precise horizontal location of the top particle.
Once the direction of motion in configuration space is chosen,
$\alpha_{s,n}$ is nearly zero during subsequent evolution.

We also solved for the full particle trajectories using the
instantaneous rigidity matrices obtained from experiments by
numerically solving Newton's equations of motion for the configuration
space vector ${\bf r}$:
\begin{equation}
\label{newton}
M\frac{d^2 {\bf r}}{dt^2} = -K {\bf r} - b \frac{d{\bf r}}{dt} 
- {\bf g},
\end{equation}
where $M$ is the mass matrix and $b$ is the damping parameter.  We
solved Eq.~\ref{newton} in the overdamped limit, {\it i.e.}, $d{\bf
r}/dt = -K {\bf r}/b -{\bf g}/b$, and find that the
simulations are able to accurately recapitulate the full particle
trajectories in the experiments as shown by the solid line in
Fig.~\ref{fig:two} (a).

We performed quasi-two dimensional experiments of granular materials
undergoing vibration-induced compaction. Using accurate particle
tracking techniques, we monitored the motion of all grains in the
system following successive vibration bursts. Using the center of mass
to classify the static packings, we find that they are organized into
geometrical families. Abrupt
changes in the direction of the evolution of the family in
configuration space signal changes in the contact network. These
experimental results represent a breakthrough in our understanding 
frictional granular packings and confirm recent simulations that identified
geometrical families in isotropically compressed frictional
packings~\cite{shen}. We also modeled each contact network using rigid
links between contacting grains and showed that grain motion between
vibration bursts occurs in the null space of the under-coordinated
rigidity matrix. More specifically, we find that the particles move in
the direction of gravity projected into the null space. This novel
method for determining particle motion can be applied to other driving
mechanisms, {\it e.g.} fixed applied shear stress, and will allow us
to understand shear-induced jamming and protocol-dependent mechanical
properties in frictional packings.

\begin{acknowledgements}
We acknowledge financial support from the W. M. Keck Foundation Grant
No. DT061314 (C.S.O.) and the National Science Foundation (NSF) Grant
No. DMR-0934206 (A.H. and M.D.S.). We also acknowledge support from
the Kavli Institute for Theoretical Physics (through NSF Grant
No. PHY-1125915), where some of this work was performed.  This work
also benefited from the facilities and staff of the Yale University
Faculty of Arts and Sciences High Performance Computing Center and the
NSF (Grant No. CNS-0821132) that in part funded acquisition of the
computational facilities.
\end{acknowledgements}

\end{document}